\documentclass[english,a4paper,twocolumn,prl,amsmath,amssymb,tightenlines,reprint]{revtex4-2}
\usepackage{bm}
\usepackage{graphicx,xcolor}
\usepackage{soul}
\definecolor{linkcolor}{HTML}{004191}
\usepackage[colorlinks,linkcolor=linkcolor,citecolor=linkcolor, urlcolor=linkcolor]{hyperref}
\usepackage{nicematrix}
\usepackage{mlmodern}

\newcommand{\pd}[2]{\frac{\partial #1}{\partial #2}}

\newcommand{\mean}[1]{\langle #1 \rangle}

\newcommand{\adj}{\mathrm{adj}}
\renewcommand{\vec}[1]{\bm{\mathbf{#1}}}
\newcommand{\mat}[1]{\bm{\mathbf{#1}}}

\newcommand{\al}{\alpha}

\newcommand{\kap}{\kappa}

\newcommand{\sig}{\sigma}
\newcommand{\om}{\omega}

\newcommand{\kT}{k_\text{B}T}

\newcommand{\W}{\vec W_{i}}
\newcommand{\kp}{k_{+I}}
\newcommand{\km}{k_{-I}}
\newcommand{\kpm}{k_{\pm I}}
\newcommand{\tup}[2]{\tau_{#1}^{/ #2}}
\newcommand{\tdown}[2]{\tau_{#1\backslash #2}}

\begin{document}

\title{Mutual Linearity is a Generic Property of Steady-State Markov Networks}

\author{Robin Bebon}
\author{Thomas Speck}
\email{thomas.speck@itp4.uni-stuttgart.de}
\affiliation{Institute for Theoretical Physics IV, University of Stuttgart, Heisenbergstr.~3, 70569 Stuttgart, Germany}

\date{\today}

\begin{abstract}
  Understanding and predicting how complex systems respond to external perturbations is a central challenge in nonequilibrium statistical physics. Here we consider continuous-time Markov networks, which we subject to perturbations along a single edge. We find that in steady state the probabilities of any two states are linearly related to one another. We show that this mutual linearity of probabilities extends to a broad class of observables, including currents but also generic counting and state-dependent observables. Moreover, we derive an exact relation between the relative response of any state's probability and the ratio of two steady-state probabilities. Leveraging the Markov chain tree theorem, we further show that probabilities and the considered observables are constrained by the topological and kinetic properties of the network and provide analytical expressions in terms of spanning tree polynomials. Our results are general, holding for arbitrary rate parameterizations and extending far from equilibrium.
\end{abstract}

\maketitle


\emph{Introduction.}---Predicting how physical systems respond to external perturbations is a longstanding challenge in physics. For small perturbations, linear response theory provides a direct connection between the basic properties of an unperturbed system, in form of its transport coefficients, and its response~\cite{kubo57, hanggi82, baroni87}. In thermal equilibrium, this response is entirely characterized by equilibrium fluctuations via the well-known fluctuation-dissipation theorem~\cite{kubo66}. More recently, the development of stochastic thermodynamics~\cite{seifert12, peliti21, seifert25} has sparked growing interest in understanding the response of nonequilibrium systems that capture the dissipative dynamics of, e.g., chemical reaction networks~\cite{rao16, gaspard04a, mcquarrie67, mou86, schmiedl07b}, or active matter~\cite{dalcengio19, caprini21a, bebon25, kirkpatrick25, davis24}. Numerous generalizations and extensions of the fluctuation-dissipation theorem have since been proposed~\cite{agarwal72, speck06, blickle07, speck09, seifert10, seifert10a, prost09, altaner16, baiesi09, dechant20, aslyamov25, ptaszynski24a, chun21, zheng24a}.

In parallel, predominantly for continuous-time Markov jump processes, a variety of approaches have been explored to characterize the nonequilibrium response of different observables. Drawing inspiration from thermodynamic uncertainty relations~\cite{barato15, horowitz17, horowitz20, dieball23}, the response of steady-state observables~\cite{owen20, fernandesmartins23}---and their precision~\cite{ptaszynski24, aslyamov25, ptaszynski25}---is fundamentally constrained by dissipation. Similarly, topological constraints encoded in the incidence matrix provide exact response relations~\cite{aslyamov24a} and structural features such as the size of the support of the perturbation~\cite{owen23} delineates the sensitivity of steady-state observables. 

A particularly intriguing response property was unveiled by Harunari et al.~\cite{harunari24} in proving that, upon perturbing a single edge's rates, any two currents are linearly related. This holds arbitrarily far from equilibrium and follows independent of thermodynamic consistency constraints, i.e., local-detailed balance, making it a fundamental property of any irreducible Markov jump process. Further works were devoted to enrich the results of Ref.~\cite{harunari24} by assigning response coefficients a clear physical interpretation in form of first-passage times~\cite{khodabandehlou25} and proving mutual linearity relations for currents to hold more generally in cases where perturbations are supported by multiple edges~\cite{dalcengio25}. In this letter, we prove an even more fundamental linear relation between any two state-steady probabilities, from which it is straightforward to show that mutual linearity holds for edge currents. We apply our result to a broad class of observables, including state dependent and generic counting observables. As a consequence, any combination of such observables, e.g., the probability to occupy state $n$ and the traffic through edge $n\leftrightarrow m$, are linearly related, and we provide two different analytical expressions for their respective susceptibilities. Additionally, as a corollary, we find that the relative response of any state in the system is identical and determined by the ratio of only two probabilities---the probabilities of the states connected via the input edge.


\emph{Setup and Results.}---We consider continuous-time Markov networks consisting of a discrete set of $|\mathcal{N}|$ vertices (or states) $n\in \mathcal{N}$ that are connected via edges. Transitions between two states $n$ and $m$ occur with transition rates $k_{nm}$ from $n\to m$. We assume networks to be irreducible, meaning that every state $n$ can be reached from any other state $m$ within the network in finite time; this also guarantees the existence of a unique stationary state~\cite{ross14}. Probability rates and the network topology are encoded in the rate matrix $\mat{W}$ with off-diagonal $[\mat{W}]_{n,m} = k_{mn}$ and diagonal entries $\left[\mat{W}\right]_{n,n} = -\sum_{m\neq n} k_{nm}$ such that columns sum to zero to ensure probability conservation. The time evolution of the probability distribution is governed by the master equation $\partial_t \vec p = \mat{W} \vec p$ and we collect occupation probabilities of the stationary solution ($\partial_t \vec p =0$) in the normalized vector $\vec p = (p_1, p_2, \ldots, p_{|\mathcal{N}|})^\top$.

\begin{figure}
  \includegraphics{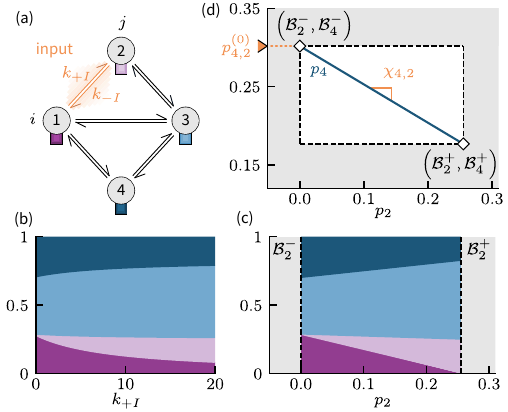}
  \caption{(a) Markov network consisting of four states. The controlled input edge $I$ (orange) is spanned by states $i = 1$ and $j = 2$. (b) Occupation probabilities as function of rate $\kp$ while $\km$ is fixed. Probabilities are colored according to the labels in (a). (c) Linear relation [Eq.~\eqref{eq:linearity}] between probabilities. All probabilities are linear in the reference probability $p_2$. Dashed lines depict the limits of values accessible to $p_2$ [Eq.~\eqref{eq:bounds}]. (d) Probability $p_4$ over $p_2$ together with the rectangular phase space region spanned by their respective bounds and markers indicate the corners of the diagonal that contains the possible solutions. The affine coefficient $p_{4,2}^{(0)}$ and susceptibility $\chi_{4,2}$ entering Eq.~\eqref{eq:linearity} are indicated in orange.}
  \label{fig:4state}
\end{figure}

Inspired by recent results~\cite{aslyamov24a,harunari24,khodabandehlou25}, we study the stationary response of Markovian jump processes to changes in the rates $\kp \equiv k_{ij}$ and $\km \equiv k_{ji}$ of an externally controlled edge $I \equiv i\leftrightarrow j$ [see Fig.~\ref{fig:4state}(a)], henceforth referred to as the input edge. In general, we allow simultaneous variation of both input rates, or each individually, while the remaining network is kept unchanged. As one of the fundamental quantities of a Markov network, we focus our attention on steady-state probabilities and leverage the relation
\begin{align}
  \partial_k \vec{p} = - \mat{W}_n^{-1} \left( \partial_k \mat{W}_n \right)\vec p,
  \label{eq:response_prob}
\end{align}
which has been derived recently and provides a linear algebra-based method to determine nonequilibrium responses~\cite{aslyamov24a}. Here $k$ denotes an arbitrary rate, or more generally rate parameter, and we define $\mat{W}_n$ as the rate matrix $\mat{W}$ with the $n$th row replaced by ones [cf.~Eq.~\eqref{app_eq:W_n}]. Unlike the Markov generator matrix, $\mat{W}_n$ is invertible for any $n$~\cite{aslyamov24,aslyamov24a,harunari24} and normalization dictates $\mat{W}_n \vec p = \vec e_n$, where $\vec e_n$ is the unit vector with one as its $n$th element. Equation~\eqref{eq:response_prob} follows from simple differentiation.

Plugging $\W$ and $k=\kpm$ into Eq.~\eqref{eq:response_prob}, we find that the responses of any two states $n$ and $m\in \mathcal{N}$ satisfy $\partial_{\kpm} p_n /(\partial_{\kpm} p_m) = \chi_{n,m}$, where $\chi_{n,m}$ is independent of rates $\kpm$. The probabilities thus satisfy 
\begin{align}
  p_n(\kpm) = p_{n,m}^{(0)} + \chi_{n,m} p_m(\kpm)
  \label{eq:linearity}
\end{align}
for all values of $\kpm$ associated with the freely chooseable input edge $I$~\footnote{Probability conservation enforces $\sum_{n=1}^{|\mathcal{N}|} p_{n,m}^{(0)} = 1$ and $\sum_{n=1}^{|\mathcal{N}|} \chi_{n,m}=0$.}. This constitutes our first main result. For a detailed proof, see \emph{Appendix A}. Although probabilities are nonlinear functions of the transition rates [cf.~Fig.~\ref{fig:4state}(b)], Eq.~\eqref{eq:linearity} states that changing the rates of a single edge triggers a linear response between any two states $n$ and $m$ [cf.~Fig.~\ref{fig:4state}(c)], with $\chi_{n,m}$ denoting the probability-probability susceptibility---i.e., the sensitivity of probability $p_n$ to changes in $p_m$. It is given by the algebraic expression
\begin{align}
  \chi_{n,m} \equiv \frac{\left[\adj\left(\W\right)\right]_{n, j}}{\left[\adj\left(\W\right)\right]_{m, j}},
  \label{eq:suscs}
\end{align}
where $\left[\adj(\mat{M})\right]_{n, m} \equiv (-1)^{n+m}\det\left(\mat M_{\backslash (m,n)}\right)$ denotes elements of the adjugate matrix, or equivalently the transpose of the cofactor matrix, of a square matrix $\mat{M}$. The subscript ``$\backslash(n,m)$'' indicates deletion of the $n$th row and $m$th column. Susceptibilities defined through Eq.~\eqref{eq:suscs} remain invariant to changes in the input rates since they appear only in the $j$th row of $\W$ [cf.~Eq.~\eqref{app_eq:W_n}]. Furthermore, they are invariant under global rescaling of rates and satisfy the identities
\begin{align}
  \chi_{n,m} = \frac{1}{\chi_{m,n}} \qquad \frac{\chi_{n,m}}{\chi_{n,l}} = \chi_{l,m} \qquad \frac{\chi_{n,m}}{\chi_{l,m}} = \chi_{n,l}.
  \label{eq:susc_relation}
\end{align}
Remarkably, a change of reference state, e.g., $m\to l$ in Eq.~\eqref{eq:suscs}, results in a global rescaling of all susceptibilities by the constant factor $1/\chi_{l,m}$. The same holds for changes in the target probability. Thus, the normalized magnitude of susceptibilities is universal within any network. If both states coincide, $\chi_{n,n} = 1$, as expected. At this stage the affine coefficients in Eq.~\eqref{eq:linearity} remain undetermined integration constants. We will derive exact expressions for these offsets later. 

As a direct corollary of Eqs.~\eqref{eq:linearity} and~\eqref{eq:suscs}, we obtain a relation between the ratio of an arbitrary state's response to changes in the forward and backward rates along the input edge and the probability of states that span that edge
\begin{align}
  -\frac{\partial_{\kp} p_n}{\partial_{\km} p_n} = \frac{p_{i}}{p_{j}}.
  \label{eq:ratio}
\end{align}
For $p_j = 0$ the response $\partial_{\km} p_n$ vanishes [Eq.~\eqref{eq:response}] and taking the ratio is mathematically undefined. From an operational perspective, Eq.~\eqref{eq:ratio} allows the direct inference of each state's relative sensitivity from the occupation probabilities of only two states; and vice versa. Moreover, if the input edge obeys detailed balance, $p_{i}\kp = p_{j}\km$, Eq.~\eqref{eq:ratio} reduces to $-\partial_{\kp} p_n/(\partial_{\km} p_n) = \km/\kp$, and links the nonequilibrium responses of all states in the system to the equilibrium thermodynamic potentials encoded in the transition rates of a single edge. Employing the results of Ref.~\cite{liang24}, we show in \cite{sm} that the response ratio in Eq.~\eqref{eq:ratio} can be bounded in terms of pseudo-equilibrium quantities that depend on thermodynamic properties only.

\emph{Probability bounds.}---Next we use standard graph theory identities to bound the region of phase space accessible to stationary probabilities when controlling a single edge's transition rates. According to Kirchhoff's Markov chain tree theorem~\cite{hill66, schnakenberg76, tutte05} the stationary probability of a given state $n$ can be expressed as $p_n=\tau_n /\sum_y \tau_y$. Here, $\tau_n = \sum_{\mu} w(\mathcal{T}^\mu_n)$ sums over all spanning trees $\mu$, and $w(\mathcal{T}^\mu_n)$ is the product of transition rates along the directed tree $\mathcal{T}_n^\mu$ rooted at $n$~\footnote{Rooted trees $\mathcal{T}^\mu_n$ are defined as directed subgraphs containing all nodes and no cycles, with every node---except the root $n$---having outdegree one.}. To isolate the influence of the external control, we introduce the decomposition~\cite{harunari24}
\begin{equation}
  \tau_n = \tau_{n\backslash i \leftrightarrow j} + \kp \tau_{n}^{/i\to j} + \km \tau_{n}^{/j\to i},
  \label{eq:split}
\end{equation}
where the first term on the right-hand side denotes the sum over the subset of trees that exclude any transitions through edge $I$. The term $\tau_{n}^{/i\to j}$ (resp.~$\tau_{n}^{/j\to i}$) represents the sum over trees that do contain transition $i\to j$ (resp.~$j\to i$) while rate $\kp$ (resp.~$\km$) is factored out of the product. The decomposition Eq.~\eqref{eq:split} is a linear function of the input rates, while residual polynomials [$\tau$'s on the right-hand-side of Eq.~\eqref{eq:split}] are independent of $\kpm$. These polynomials are easily evaluated numerically~\cite{sm}. 

Using the Markov chain tree theorem together with Eq.~\eqref{eq:split}, we can formally evaluate the limits $\lim_{\kpm\to \infty}p_n$ to obtain the (asymptotically exact) bounds 
\begin{align}
  \mathcal{B}^+_n \equiv  \frac{\tau_n^{/i\to j}}{\sum_l \tau_l^{/i\to j}} \qquad \mathcal{B}^-_n\equiv \frac{\tau_n^{/j \to i}}{\sum_l \tau_l^{/j\to i}},
  \label{eq:bounds}
\end{align}
that limit the achievable values of $p_n$. Due to the non-monotonic response of probabilities with respect to input rates [cf.~Fig.~\ref{fig:4state}(b)], determining which of the bounds from Eq.~\eqref{eq:bounds} corresponds to the upper [$\max(\mathcal{B}^+_n,\mathcal{B}^-_n)$] and lower boundary [$\min(\mathcal{B}^+_n,\mathcal{B}^-_n)$] of solution space, must be done \emph{a posteriori} and separately for each state $n$~\footnote{For a concrete example, consider Fig.~\ref{fig:4state}(d). Here, the upper $\mathcal{B}^+_2$ ($\mathcal{B}^-_4$) and lower $\mathcal{B}^-_2$ ($\mathcal{B}^+_4$) bounds of $p_2$ and $p_4$ are mirrored due to their different scalings  with $k_{+I}$ [cf.~Fig.~\ref{fig:4state}(b)].}. 

For any pair of states $n$ and $m$, the bounds [Eq.~\eqref{eq:bounds}] span a (rectangular) region in the $(p_n,p_m)$-plane that contains all steady-state solutions achievable through changes in input edge $I$. Invoking the mutual linearity condition from Eq.~\eqref{eq:linearity}, we know \emph{a priori} that all possible solution pairs $(p_n,p_m)$ condense onto a straight line and thus all permitted solutions lie on the diagonal spanned between $(\mathcal{B}^-_n, \mathcal{B}^-_m)$ and $(\mathcal{B}^+_n, \mathcal{B}^+_m)$; the corners of the rectangular phase space region [cf.~Fig.~\ref{fig:4state}(d)]. The offset and slope of the diagonal provide explicit expressions for the affine coefficient and susceptibility of Eq.~\eqref{eq:linearity}
\begin{align}
  p_{n,m}^{(0)} = \frac{\mathcal{B}^-_n\mathcal{B}^+_m - \mathcal{B}^-_m \mathcal{B}^+_n}{\mathcal{B}^+_m-\mathcal{B}^-_m} \qquad
  \chi_{n,m} = \frac{\mathcal{B}^+_n-\mathcal{B}^-_n}{\mathcal{B}^+_m-\mathcal{B}^-_m}.
  \label{eq:offset_suscs}
\end{align}
The latter presents an alternative to the linear algebra-based susceptibilities from Eq.~\eqref{eq:suscs} in terms of spanning tree ensembles. Note that $\mathcal{B}^+_{i} = 0$ since transitions $i\to j$ are prohibited for any tree rooted in $i$ and hence $\tau_{i}^{/ i \to j} = 0$; the same holds for $\mathcal{B}^-_{j}$. As a result, the offset in Eq.~\eqref{eq:offset_suscs} simplifies to, e.g., $p_{n,j}^{(0)} = \mathcal{B}^-_{n}$ [cf.~Fig.~\ref{fig:4state}]. 

In \emph{Appendix B} we provide additional bounds for the case of single rate control or unidirectional input edges. The corresponding susceptibilities and offsets follow in analogy to Eq.~\eqref{eq:offset_suscs}.

\emph{General observables.}---We now shift our attention to a broad class of observables that we express as
\begin{align}
  \mathcal{O}_A \equiv  \al \sum_{m\in A} c_m p_m + \beta \sum_{\substack{l,m \in A \backslash \lbrace I \rbrace}} d_{ml} k_{ml} p_m
  \label{eq:obs}
\end{align}
with constant coefficients $c_m$ and $d_{mn}$. The sums run over all states and transitions contained in $A \subseteq \mathcal{N}$, excluding $I$. For $\beta = 0$, Eq.~\eqref{eq:obs} describes a state dependent observable that weighs state occupation with weights $c_m$. Conversely, $\al=0$ defines a counting observable that is either traffic or current-like depending on the signs of increments $d_{ml}$. In case $d_{ml} = d_{lm}$, Eq.~\eqref{eq:obs} is time-symmetric and, for unit coefficients, reduces to the dynamical activity (rate)~\cite{merolle05, maes17a, maes17, diterlizzi19} (closely related to frenesy~\cite{maes20a}) of all edges supported by the states of $A\backslash \lbrace I \rbrace$. Similarly, for $d_{ml} = -d_{lm} = 1$, Eq.~\eqref{eq:obs} measures directed flows and reduces to the edge current $j_{ml}$ if $A$ encompasses only states $m$ and $l$. 

As a straight forward generalization of Eq.~\eqref{eq:linearity}, we find that all observables of the form in Eq.~\eqref{eq:obs} are likewise linear to each other and satisfy
\begin{align}
  \mathcal{O}_A = \left(\mathcal{O}_{A,n}^{(0)} - \frac{\mathcal{X}_{A,n}}{\mathcal{X}_{B,n}} \mathcal{O}_{B,n}^{(0)} \right) + \frac{\mathcal{X}_{A,n}}{\mathcal{X}_{B,n}} \mathcal{O}_B,
  \label{eq:obs_lin}
\end{align}
with 
\begin{align}
  \begin{bmatrix}  
  \mathcal{O}_{A,n}^{(0)} \\[.7ex]
  \mathcal{X}_{A,n} \\[.7ex]
  \mathcal{B}^\pm_{\mathcal{O}_A}
  \end{bmatrix} 
  &\equiv  \al \sum_{m\in A} c_m   
  \begin{bmatrix} 
    p_{m,n}^{(0)} \\[.7ex] 
    \chi_{m,n} \\[.7ex]
    \mathcal{B}^\pm_n
  \end{bmatrix} + \beta \sum_{\substack{l,m \in A \backslash \lbrace I \rbrace}} d_{ml} k_{ml} 
  \begin{bmatrix} 
    p_{m,n}^{(0)} \\[.7ex] 
    \chi_{m,n} \\[.7ex]
    \mathcal{B}^\pm_n
  \end{bmatrix}.
  \label{eq:offset_sucs_bounds}
\end{align}
The last line defines the generalization of the bounds from Eq.~\eqref{eq:bounds}, which can be used to determine mesoscopic susceptibilities $\mathcal{X}_{A,n}$ and $\mathcal{O}^{(0)}_{A,n}$ through Eq.~\eqref{eq:offset_suscs}. In case $\mathcal{O}_B = 0$ for all $\kpm$, i.e., $\mathcal{B}^\pm_{\mathcal{O}_B} = 0$, Eq.~\eqref{eq:obs_lin} no longer holds since susceptibilities are mathematically undefined~\footnote{One example are currents through bridges, which are zero regardless of input rates.}.

For single-rate perturbations and time-asymmetric counting observables on edge~$I$, i.e., $d_{ij} = -d_{ji}$, the second summation in Eq.~\eqref{eq:obs} can be extended to include edge~$I$, as the explicit dependence on the rates~$\kpm$ cancels~\cite{sm}. This allows us to immediately recover the key results of Refs.~\cite{harunari24,khodabandehlou25}, where the linearity of currents was derived by other means. Our findings augments these works by proving that mutual linearity is a fundamental property of steady-state probabilities [Eq.~\eqref{eq:linearity}], from which it naturally extends to a broad class of observables [see Eqs.~\eqref{eq:obs} and~\eqref{eq:obs_lin}], including currents.

\begin{figure*}
  \includegraphics{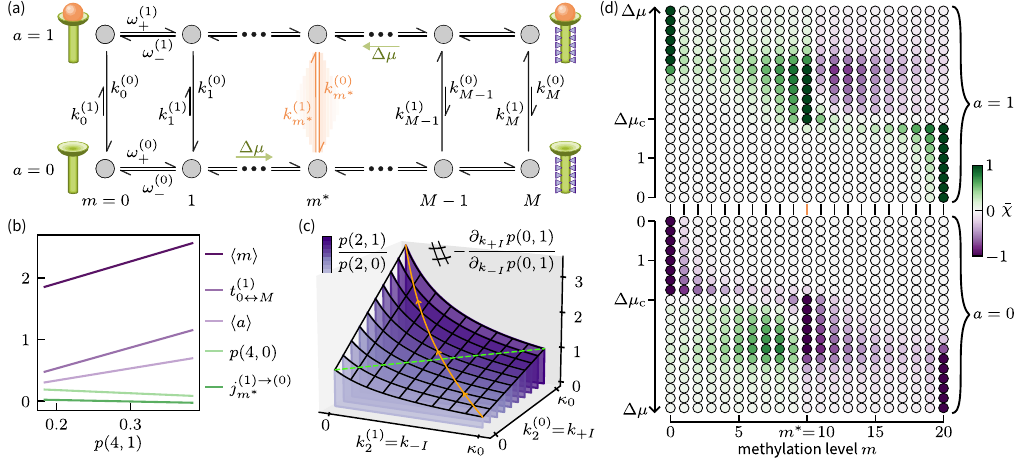}
  \caption{(a)~Adaptation model for a chemoreceptor with a total of $|\mathcal{N}|=2(M+1)$ states comprising the activity state $a\in\lbrace 0,1\rbrace$ and the methylation level $m \in \lbrace 0, \ldots, M \rbrace$ (purple triangles). Occupation probabilities are denoted as $p(m,a)$. (b)~Mutual linearity between the average methylation $\langle m \rangle$ [Eq.~\eqref{app_eq:avgm}], average activity $\langle a \rangle$ [Eq.~\eqref{app_eq:ptumble}], mesoscopic edge traffic $t^{(1)}_{0\leftrightarrow M}$ [Eq.~\eqref{app_eq:traf}], and current through the input edge $j_{m^\ast}^{(1)\to (0)}$ [Eq.~\eqref{app_eq:cur}], and probabilities $p(4,0)$ and $p(4,1)$. (c)~Comparison between the ratio of probabilities connected by the input edge $p(2,1)/p(2,0)$ (colors) and the relative response $-\partial_{\kp} p(0,1)/[\partial_{\km} p(0,1)]$ (wireframe). For the dashed green line, edge $I$ obeys detailed balance and the probability and response ratios equal $\km/\kp$. The solid orange line indicates the protocol we adopt for input rates in panels (b) and (d). In panels (b) and (c) we consider a sensory network with $M=4$ methylation levels, $\Delta \mu=1.3$, and the input edge connects activity states at methylation level $m^\ast=2$. (d)~Normalized susceptibilities of the adaptation model with $M=20$ methylation levels as we vary the driving strength $\Delta \mu$. Each circle corresponds to a state of the active $a=1$ (top) or passive $a=0$ branch (bottom) and is colored according to its susceptibility obtained from normalizing the results of Eq.~\eqref{eq:suscs}, with reference probability $p(10,1)$. The input edge is located at $m^\ast = 10$ and indicated in orange.}
  \label{fig:chemotaxis}
\end{figure*}

\emph{Sensory adaptation.}---To illustrate the results of this letter, we adopt an idealized biophysical model of sensory adaptation that is motivated by the chemotaxis pathway in \emph{E.~coli}~\cite{tu08, lan12, murugan17}. The process is described through the Markovian network shown in Fig.~\ref{fig:chemotaxis}(a) and captures the mesoscopic dynamics of the binary activity state $a\in \lbrace 0,1\rbrace$ and the degree of methylation $m$ of a concentration sensing chemoreceptor~\cite{barkai97, tu13}.
We assume that the transition rates between the activity states at $m=m^\ast$ are sensitive to an external signal and we treat it as input edge in the following. For chemotaxis adaptation, this signal relates to the (logarithmic) concentration of chemoattractants~\cite{endres06,kalinin09, skoge13}. The remaining rates between activity states are left constant and biased according to the arrows in Fig.~\ref{fig:chemotaxis}(a). For given activity, transitions are independent of the degree of methylation and the system relaxes into the energetically favored state of low (high) methylation for the inactive (active) branch, unless the chemical energy input $\Delta \mu$ is sufficiently strong to drive transitions uphill the free energy landscape [cf.~Fig.~\ref{fig:probabilities_chemo} in \emph{Appendix D}]. Further details are provided in the Supplementary Material~\cite{sm}. 

In Fig.~\ref{fig:chemotaxis}(b) and (c), we illustrate our main results from Eqs.~\eqref{eq:linearity},~\eqref{eq:obs_lin}, and~\eqref{eq:ratio} for an adaptation ladder containing $M=4$ methylation levels and an input edge connecting activity states at $m^\ast=2$. Figure~\ref{fig:chemotaxis}(b) depicts the linear relationship between various state observables ($\langle a \rangle$, $\langle m \rangle$), counting observables ($j_{m^\ast}^{(1)\to (0)}$, $t_{0\leftrightarrow M}^{(1)}$) and probabilities $p(m,a)$. Susceptibilities and offsets follow from Eqs.~\eqref{eq:suscs},~\eqref{eq:offset_suscs}, and~\eqref{eq:offset_sucs_bounds} respectively.    
For the relative response of, e.g., probability $p(0,1)$, we confirm Eq.~\eqref{eq:ratio} in Fig.~\ref{fig:chemotaxis}(c), which thus enables the inference of the networks sensitivity to changes in rates $\kpm$ through limited knowledge in form of the two probabilities $p(2, \lbrace 0,1\rbrace)$. Specifically, for $\kp > \km$, probabilities are more responsive to changes in $\kp$ and vice versa. For $\kp = \km$ probabilities are equally responsive which, owing to the symmetric choice of transition time scales along activity axes, coincides with a vanishing current through the controlled edge. 

Depending on the chemical energy input $\Delta \mu$ the system operates in either of two different modes: sensitive or adaptive~\cite{kharbanda24}. For $\Delta \mu < \Delta \mu_c$ the system's activity is highly sensitive to alterations in the external signal, biasing relaxation into either the active $p(M,1)$ or inactive $p(0,0)$ state. Conversely, beyond a critical driving strength $\Delta \mu > \Delta \mu_\mathrm{c}$ the system compensates changes in the external signal by adapting its internal methylation state and ensures a robust activity output $\langle a \rangle$ by stabilizing dissipative cycles around the adapted level of methylation (here $m^\ast$ for $S=0$)~\cite{sm}. Indeed, Ref.~\cite{sartori15} shows that the onset of adaptation for such models can be described as a (continuous) nonequilibrium phase transition with $\Delta \mu$ acting as control parameter. 

Employing Eq.~\eqref{eq:linearity} to an extended system with $M=20$ levels, we find in Fig.~\ref{fig:chemotaxis}(d) that the adaptation onset coincides with a pronounced change in the susceptibilities. Notably, the depicted normalized susceptibilities are, apart from their sign, independent of the chosen reference state [cf.~Eq.~\eqref{eq:susc_relation}]. The susceptibility patterns are most easily understood by considering how probability is redistributed within the system [cf.~Fig.~\ref{fig:probabilities_chemo}]: For $\Delta \mu < \Delta \mu_\mathrm{c}$, probability is concentrated in the energetically favored states $p(0,0)$ and $p(M,1)$. Because these states are highly populated, even small perturbations in the input produce comparatively large changes in their stationary weights, rendering them most susceptible. Conversely, if the critical driving strength is exceeded, input-induced changes predominantly redistribute probability among the states that participate in dissipatively stabilized cycles, causing susceptibilities to localize around the adapted methylation level $m^\ast$. The number of methylation levels traversed by these cycles increases with $\Delta \mu$ until probability is collected in $p(M,0)$ and $p(0,1)$ through persistent edge currents.

\emph{Discussion.}---In this letter, we prove that any two steady-state probabilities of continuous-time Markov networks are linearly related [Eq.~\eqref{eq:linearity}]. We show that this linearity relation translates to a broad class of observables [Eq.~\eqref{eq:obs_lin}], including state-dependent and counting observables, which allows us to recover and considerably extend earlier results on the mutual linearity of steady-state currents reported in Refs.~\cite{harunari24,khodabandehlou25}. 

All coefficients entering our main results in Eqs.~\eqref{eq:linearity} and~\eqref{eq:obs_lin} are empirically accessible through measuring two combinations of probabilities (e.g.~through fluorescence techniques~\cite{tietz06, chung12,chung13}, force~\cite{stigler11, petrosyan21} or plasmon ruler spectroscopy~\cite{ye18, vollmar24}, etc.) or observables under different external input. Hence, additional measurements have predictive power and comparing susceptibilities offers a robust method for proof testing viable candidate models~\cite{harunari24}. Moreover, once coefficients of affine transformations are established (e.g., between a current and the total dynamical activity), measurement of either of the two quantities allows simultaneous inference of the other. We present two methods to numerically evaluate susceptibilities: (i) Through the linear-algebra based expression in Eq.~\eqref{eq:suscs}, or (ii) by evaluating tree polynomials that enter the bounds in Eq.~\eqref{eq:bounds} [resp.~Eq.~\eqref{eq:offset_sucs_bounds}]. These bounds additionally offer analytical expressions for the affine coefficients entering Eqs.~\eqref{eq:linearity} and~\eqref{eq:obs_lin} and delineate the space of possible solutions.

Our results are general, hold arbitrarily far from thermal equilibrium, and make no assumptions about the parametrization of transition rates. If instead one assumes an explicit parametrization in form of Arrhenius~\cite{arrhenius89} or Eyring-like~\cite{eyring35} rates, we prove a set of response relations for observables $\mathcal{O}_A$ [Eq.~\eqref{eq:obs}] in Ref.~\cite{sm} that encompass previous identities~\cite{aslyamov24} upon choosing the corresponding observables, but also hold more broadly for nonlocal responses. Furthermore, edge transitions are not required to be bidirectional (see~\emph{Appendix C}) and transitions over an edge can occur through multiple reaction channels. 

Throughout, we focused on continuous-time Markov networks and leverages the properties of the rate matrix $\vec W$ to derive Eq.~\eqref{eq:linearity} and corollary results. Since similar type of matrices are encountered in various other contexts, e.g., for the voter model~\cite{clifford73, sood05, angelesserrano09, masuda09}, it would be interesting to explore the ramifications of our work also in these systems.

\begin{acknowledgments}
	\emph{Acknowledgments.}---We acknowledge financial support by the Deutsche Forschungsgemeinschaft (DFG) through the collaborative research centers TRR 146 (grant no. 233630050) and SFB 1551 (grant no. 464588647).
\end{acknowledgments}

%

\onecolumngrid
\section*{End Matter}
\twocolumngrid
\renewcommand{\theequation}{A\arabic{equation}}
\setcounter{equation}{0} 

\emph{Appendix A: Proof of linearity of steady-state probabilities.}---We use Eq.~\eqref{eq:response_prob} to calculate the response of steady-state probabilities subjected to perturbations in the rates $\kpm$. We consider the matrix
\begin{align}
 \mat{W}_{i} = \begin{pNiceMatrix}[last-col,first-row]
  & i  & j  & & \\ 
  \ddots& &  & &\\
  & 1 & 1 & & i\\[1ex]
  & \kp & -\km-\ldots & & j  \\
       &  &  & \ddots & \\
\end{pNiceMatrix}
\label{app_eq:W_n}
\end{align}
in which the appearance of input rates $\kpm$ is reduced to only two entries: $(i,j)$ and the diagonal element $(j,j)$. We find 
\begin{align}
  \partial_{\kp} p_n &= - \left[\mat{W}^{-1}_{i}\right]_{n,j} p_{i} = - \frac{\left[\adj\left(\mat{W}_{i}\right)\right]_{n,j}}{\det\left( \mat{W}_{i}\right)}  p_{i} \nonumber \\ 
  \partial_{\km} p_n &=  \left[\mat{W}^{-1}_{i}\right]_{n,j} p_{j} =  \frac{\left[\adj\left(\mat{W}_{i}\right)\right]_{n,j}}{\det\left( \mat{W}_{i}\right)}  p_{j}
  \label{eq:response}
\end{align}
for an arbitrary vertex $n$. The second equalities are obtained by exploiting the relation between adjugate matrix $\adj\left(\mat{W}_{i}\right)$, with $\left[\adj(\W)\right]_{n, m} \equiv (-1)^{n+m}\det\left(\mat{W}_{i\backslash (m,n)}\right)$---this is equivalent to the transpose of the cofactor matrix of $\W$---and inverse of $\mat{W}_i$
\begin{align}
  \adj\left(\mat{W}_{i}\right) =  \mat{W}^{-1}_{i}\det\left( \mat{W}_{i}\right).
  \label{eq:adj}
\end{align}
As our main results, we find from Eqs.~\eqref{eq:response} and~\eqref{eq:adj}
\begin{align}
  \frac{\partial_{\kpm} p_n}{\partial_{\kpm} p_m} = \frac{\left[\adj\left(\W\right)\right]_{n, j}}{\left[\adj\left(\W\right)\right]_{m, j}} \equiv \chi_{n,m}.
\end{align}
The mutual linearity relation in Eq.~\eqref{eq:linearity} then follows immediately. The relation holds except if $p_m = 0$ for all $\kpm$, which is generally impossible for irreducible graphs unless transition rates are infinitely large.

\emph{Appendix B: Probability bounds for unidirectional control.}---While the bounds in Eq.~\eqref{eq:bounds} also hold if only one of the input rates is controlled or if the edge is totally irreversible, the bound corresponding to the fixed (or zero) rate overestimates the range of accessible values. Instead, $\lim_{\kpm \to 0} p_n$ results in the bounds
\begin{align}
  \mathcal{U}^+_n \equiv  \frac{\tau_{n\backslash i \to j} }{\sum_l \tau_{l \backslash i\to j}} \qquad \mathcal{U}^-_n \equiv \frac{\tau_{n\backslash j\to i} }{\sum_l \tau_{l \backslash j \to i}},
  \label{eq:tight_bounds}
\end{align}
that  limit all possible solutions. Note that writing $\tau_{n\backslash i\to j} = \tau_{n\backslash i\leftrightarrow j} + \km \tau_{n}^{/ j\to i}$, and analog for $\tau_{n\backslash j\to i}$, the bounds are independent of the controlled rate. Analytical expressions for the affine coefficients and susceptibilities follow from Eq.~\eqref{eq:offset_suscs} upon replacing $\mathcal{B}_{\lbrace n,m\rbrace}^\pm \to \mathcal{U}_{\lbrace n,m\rbrace}^\mp$ for the respective cases.

\begin{figure}[b!]
  \includegraphics{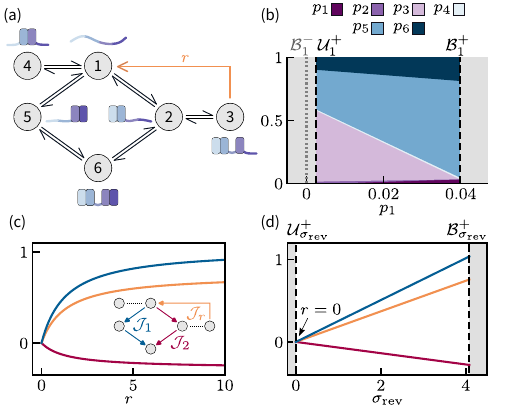}
  \caption{(a) Calmodulin folding network~\cite{stigler11} with an added unidirectional resetting transition with rate $r$ (orange) from an intermediate ``$3$'' to the unfolded state ``$1$''. The unidirectional rate is subject to external control. (b) Linear relation between steady-state probabilities and $p_1$. The black dashed lines and grey shaded areas indicate the bounds on $p_1$ given by Eqs.~\eqref{eq:bounds} and~\eqref{eq:tight_bounds} respectively. The grey dotted line shows the bound $\mathcal{B}_1^-$ [Eq.~\eqref{eq:bounds}]. (c) Currents through the two different folding pathways $\mathcal{J}_1 \equiv j_{15} + j_{56}$ and $\mathcal{J}_2\equiv j_{12}+j_{26}$, as well as the resetting flux $\mathcal{J}_r \equiv p_3r$ (see inset) over the resetting rate $r$. (d) The same observables as in (c) plotted over the entropy production rate of all microscopically reversible transitions $\sig_\mathrm{rev} \equiv \sum_{n < m\in \mathcal{N}\backslash \lbrace I \rbrace} j_{nm} \ln(k_{nm}/k_{mn})$.}
  \label{fig:calmodulin}
\end{figure}

\emph{Appendix C: Calmodulin folding.}---For a second concrete illustration, we consider an experimentally inferred folding network of single calmodulin proteins~\cite{stigler11}, which we augment by a unidirectional irreversible transition that stochastically ``resets''~\cite{evans11, fuchs16, pal17, evans20} the protein from the intermediate state ``$3$'' to the unfolded state ``$1$'' with rate $\kp \equiv r$ [Fig.~\ref{fig:calmodulin}(a)]. This unidirectional transition is subjected to external control while the remaining rates are fixed and satisfy detailed balance~\cite{sm, bebon23}, i.e., for $r=0$ the system is in thermal equilibrium. Importantly, the irreducibility assumption entering Eq.~\eqref{eq:linearity} is satisfied by networks containing unidirectional transitions, provided absorbing states are absent and all bridges---edges that connect otherwise isolated subsets of states---are microscopically reversible. 

While the probabilities to observe different protein configurations are nonlinear functions of $r$, they are linearly related to each other [Eq.~\eqref{eq:linearity}]. For instance, all probabilities vary linearly with $p_1$---the probability of the unfolded state---as depicted in Fig.~\ref{fig:calmodulin}(b). However, although we allow $r\geq 0$ to take arbitrarily large values, the redistribution of probability among states is intrinsically limited by the topology and kinetic properties of the network, which restricts viable solutions to the values bounded by $\mathcal{B}_n^+$ [Eq.~\eqref{eq:bounds}] and $\mathcal{U}_n^+$ [Eq.~\eqref{eq:tight_bounds}]. The latter replaces $\mathcal{B}^-_n$ for the unidirectional input edge considered here (see \emph{Appendix B}). 

Along similar lines, observables expressed through Eq.~\eqref{eq:obs} depend non-trivially on the controlled rate $r$ [Fig.~\eqref{fig:calmodulin}(c)]. Among themselves, however, any two observables are linearly related according to Eq.~\eqref{eq:obs_lin}. In Fig.~\eqref{fig:calmodulin}(d) we illustrate the linear relationship between the mesoscopic currents $\mathcal{J}_ {\lbrace 1, 2 \rbrace}$ through different folding pathways, the resetting flux $\mathcal{J}_r$ (see inset), and the entropy production rate of all microscopically reversible transitions $\sig_\mathrm{rev}$~\cite{pal21, pal21a}. The associated affine coefficients, susceptibilities, and bounds follow straightforwardly from Eq.~\eqref{eq:offset_sucs_bounds}.

\emph{Appendix D: Sensory adaptation.}---To illustrate the broad range of applicability of the linearity relations in Eqs.~\eqref{eq:linearity} and~\eqref{eq:obs_lin}, we examine various observables in Fig.~\ref{fig:chemotaxis}(b). As state observables we consider the average methylation level, defined as
\begin{align}
  \mean{m} \equiv \sum_{m=0}^M  m \left[p(m,0) +p(m,1) \right]
  \label{app_eq:avgm}
\end{align}
and the average output activity
\begin{align}
  \langle a \rangle \equiv \sum_{m=0}^{M}  p(m,1),
  \label{app_eq:ptumble}
\end{align}
that characterized the fraction of time the receptor spends in its active, $a=1$, state.
As counting observables we consider the mesoscopic traffic along the activity axis $a=1$
\begin{align}
t_{0\leftrightarrow M}^{(1)} &\equiv \sum_{m=0}^{M-1} \left[p(m,1) \om_+^{(0)} + p(m+1,1) \om_-^{(0)}\right],
\label{app_eq:traf}
\end{align}
and the current through the input edge at $m^\ast=2$
\begin{align}
j^{(1)\rightarrow (0)}_{m^\ast} &\equiv p(m^\ast,1) k_{m^\ast}^{(0)} - p(m^\ast,0)k_{m^\ast}^{(1)}. \label{app_eq:cur}
\end{align}
For a proof that the input current indeed satisfies a linearity relation, although it appears to depend on $\kpm$ explicitly, see the Supplemental Material~\cite{sm}.

\begin{figure}
  \includegraphics{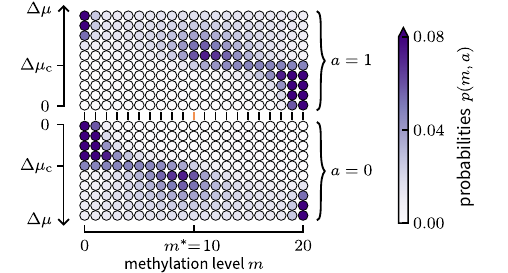}
  \caption{Steady-state probabilities of a chemotaxis ladder with $M=20$ methylation levels for various driving strengths $\Delta \mu$ (in steps of $0.5$). Each circle corresponds to a state of the active $a=1$ (top) and passive $a=0$ (bottom) branch. Colors indicate the occupation probabilities, where we cap the colorbar at a value of $0.08$ to improve readability. The input edge located at $m^\ast = 10$ is highlighted in orange. Here, we use $S=0$, such that cycles are centered around $m^\ast$~\cite{sm}.}
  \label{fig:probabilities_chemo}
\end{figure}

To corroborate the discussion of the sensory adaptation model, Fig.~\ref{fig:probabilities_chemo} depicts the steady-state probabilities of a network with $M=20$ methylation levels. The same model is considered in Fig.~\ref{fig:chemotaxis}(d).

\clearpage
\newpage
\onecolumngrid
\renewcommand{\thesection}{S\arabic{section}}
\renewcommand{\thefigure}{S\arabic{figure}}
\renewcommand{\theequation}{S\arabic{equation}}
\setcounter{equation}{0}
\setcounter{figure}{0}
\setcounter{page}{1}
\setcounter{section}{0}

\begin{center}\textbf{{\small Supplemental Material for} \\ Mutual Linearity is a Generic Property of Steady-State Markov Networks}\\[0.2cm]
Robin Bebon and Thomas Speck\\
\emph{Institute for Theoretical Physics IV, University of Stuttgart, Heisenbergstr.~3, 70569 Stuttgart, Germany}\\[0.6cm]\end{center}

\section{Mutual linearity for observables containing the input edge}
In Eq.~(9) of the main text we define a general class of observables that display linear relationships among themselves. However, the summation over transitions excludes those through edge $I$ to circumvent explicit dependencies on the perturbed, and therefore non-constant, rates $\kpm$. In this section, we show that this restriction can be lifted for two commonly encountered cases: (i) For time-asymmetric counting observables through edge $I$, i.e., $d_{ij} = -d_{ji}$, and (ii) if only a single transition rate is perturbed. The former holds for all current-like observables and the latter is ensured for unidirectional input edges, or if only one of the rates $\kpm$ is independent of the external control (e.g., when modeling the binding $\leftrightarrow$ unbinding dynamics of receptors or enzymes in response to changes in solute concentration).  

To achieve our goal, it suffices to show that a counting observable on the subset of states $\mathcal{I} = \lbrace i,j \rbrace$ satisfies a linear affine relation with the probability $p_n$ of a given state $n$. Generalization to other observables follows by virtue of Eqs.~(2) and~(10) of the main text. Thus, we consider 
\begin{align}
\mathcal{O}_\mathcal{I} = d_{ij} \kp p_i + d_{ji} \km p_j = d_{ij} \kp \left( p_{i,n}^{(0)} + \chi_{i,n} p_n  \right)  +  d_{ji} \km \left( p_{j,n}^{(0)} + \chi_{j,n} p_n  \right) \overset{?}{=} \mathcal{O}^{(0)}_{\mathcal{I}, n} + \mathcal{X}_{\mathcal{I},n} p_n
\label{eq_SI:obs}
\end{align}
and determine the conditions under which the last equality holds. For the second equality we use Eq.~(2) of the main text and, without loss of generality, take the multiplicative factor $\beta=1$. Employing Eq.~(8) together with Eq.~(7) of the main text, we express 
\begin{align}
  p_{i,n}^{(0)} = \frac{\tup{i}{j \to i} \tup{n}{i \to j}}{\Omega} \qquad \text{and} \qquad \chi_{i,n} =- \frac{\tup{i}{j\to i} \sum_l \tup{l}{i\to j} } {\Omega},
\end{align}
and analogous for $j$. Both share the normalization factor $\Omega \equiv (\mathcal{B}_n^+ - \mathcal{B}_n^-) \sum_{l} \tup{l}{i\to j} \sum_m \tup{m}{j \to i}$ and we realize $\tup{i}{i\to j} = \tup{j}{j\to i} = 0$, since per definition of rooted trees the outdegree of the root state is zero. With that, terms containing susceptibilities in Eq.~\eqref{eq_SI:obs} can be expressed as 
\begin{align}
  d_{ij} \kp \chi_{i,n} + d_{ji} \km \chi_{j,n}  = \frac{1}{\Omega} \left[ d_{ij} \sum_l \left( \tdown{l}{i \leftrightarrow j} - \tau_l \right) +  (d_{ij}+ d_{ji}) \km \sum_l \tup{l}{j\to i} \right] \tup{i}{j \to i},
  \label{eq_SI:susc}
\end{align}
where we further used the fact that $\tup{j}{i\to j} = \tup{i}{j\to i}$, alongside the decomposition from Eq.~(6) of the main text. In a similar fashion, the remaining terms of Eq.~\eqref{eq_SI:obs} can be rewritten as
\begin{align}
  d_{ij} \kp p_{i,n}^{(0)} + d_{ji} \km p_{j,n}^{(0)} = \frac{1}{\Omega} \left[ d_{ij} \sum_l \tau_l p_n - d_{ij} \tdown{n}{i\leftrightarrow j} - (d_{ij} + d_{ji}) \km \tup{n}{j\to i}\right] \tup{i}{j\to i}.
  \label{eq_SI:offset}
\end{align}
Identification of $p_n$ in the first term follows by multiplying with $\sum_l \tau_l / (\sum_l \tau_l)$ and the Markov chain tree theorem. Note that the explicit dependence on $\km$ in Eqs.~\eqref{eq_SI:susc} and~\eqref{eq_SI:offset} can be expressed in terms of rate $\kp$ if required. 

Collecting terms, Eq.~\eqref{eq_SI:obs} can be written in the desired form
\begin{align}
  \mathcal{O}_\mathcal{I} &= \underbrace{\frac{ \tup{i}{j\to i}}{\Omega} \left[- d_{ij} \tdown{n}{i\leftrightarrow j}  - (d_{ij}+d_{ji}) \km \tup{n}{j\to i} \right]}_{\equiv \mathcal{O}^{(0)}_{\mathcal{I},n}} + \underbrace{\frac{ \tup{i}{j\to i}}{\Omega} \left[ d_{ij} \sum_l \tdown{l}{i \leftrightarrow j} + (d_{ij} +d_{ji}) \km \sum_l \tup{l}{j\to i}  \right]}_{\equiv \mathcal{X}_{\mathcal{I},n}} p_n, 
\end{align}
which defines the affine coefficient $\mathcal{O}^{(0)}_{\mathcal{I},n}$ and susceptibility $\mathcal{X}_{\mathcal{I},n}$. Notably, both are constant in case a single rate (here $\kp$) is perturbed, if the input edge is unidirectional ($\km = 0$), or if increments $d_{ij} = -d_{ji}$ are asymmetric.

\section{Thermodynamic bounds on the relative response of steady-state probabilities}

\begin{figure}
  \centering
  \includegraphics{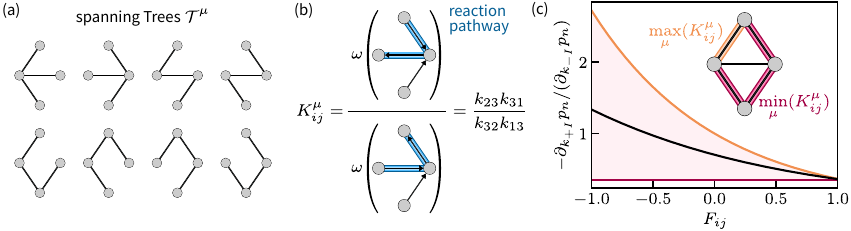}
  \caption{We revisit the 4-state model from Fig.~1(a) of the main text. (a)~Spanning trees $\mathcal{T^\mu}$ of the network. (b)~Pictorial calculation of $K_{ij}^\mu$ for an example tree. Taking the ratio of rate polynomials leaves an unique reaction pathway (blue). (c)~Upper (orange) and lower (red) bounds on the relative response of steady-state probabilities (black). The pathways that correspond to the respective bound are indicated in the inset.}
  \label{fig:4state_LDB}
\end{figure}

To assign trees $\mathcal{T}^\mu$ with a physically meaningful measure of relevance, the authors of Ref.~\cite{liang24} introduce the quantity $K_{nm}^\mu \equiv \om(\mathcal{T}_n^\mu)/ \om(\mathcal{T}_m^\mu)$. Given two states $n$ and $m$, each tree $\mathcal{T}^\mu$ defines a unique reaction pathway connecting the two and $K_{nm}^\mu$ corresponds to the ratio of products between forward and backward rates along edges of the pathway. Rates of transitions that are not contained in the reaction pathway cancel. For an illustration see Fig.~\ref{fig:4state_LDB}(a) and (b). If transition rates obey local detailed balance, $K_{ij}^\mu$ is a pseudo-equilibrium quantity that is solely determined by the thermodynamic affinity associated with specific reaction pathways in the system, which are shown to upper and lower bound the ratio of steady state probabilities~\cite{liang24} (see also Refs.~\cite{maes13, cetiner22})
\begin{align}
  \min_\mu \left( K_{nm}^\mu\right) \leq \frac{p_n}{p_m} \leq \max_\mu\left(K_{nm}^\mu\right).
  \label{eq_SI:bounds_prob}
\end{align}
Thus, the probability ratio of any two states is upper (lower) bounded by the maximal (minimal) affinity of all reaction pathways that connect these two states.

Leveraging Eq.~(5) of the main text, we immediately find the spin-off relation
\begin{align}
  \min_\mu \left( K_{ij}^\mu\right) \leq -\frac{\partial_{\kp} p_n}{\partial_{\km} p_n} \leq \max_\mu\left(K_{ij}^\mu\right),
  \label{eq_SI:TD_bounds}
\end{align}
upon substituting the probabilities of states $i$ and $j$ that span input $I$ into Eq.~\eqref{eq_SI:bounds_prob}. Intriguingly, Eq.~\eqref{eq_SI:TD_bounds} reveals that the sensitivity of probabilities is fundamentally bounded by the affinity of pathways connecting the states that span the input edge. In case of perturbations around thermal equilibrium, Eq.~\eqref{eq_SI:TD_bounds} saturates as both the upper and lower bound reduce to the Boltzmann factor.

To illustrate Eq.~\eqref{eq_SI:TD_bounds}, we revisit the 4-state model from Fig.~1(a) of the main text in Fig.~\ref{fig:4state_LDB}(c). Here, rates are parameterized such that local detailed balance is obeyed (for details see Sec.~\ref{sec:4stateLDB}). For the considered parameter range, the lower bound corresponds to the three step pathway ($1 \leftrightarrow 4 \leftrightarrow 3 \leftrightarrow 2$), whereas the upper bound---and consequently the maximal affinity---follows from transitions through the input edge ($1\leftrightarrow 2$) itself. 

\section{Specific rate parametrization and response relations}
Throughout the main text, no assumptions were made about the specific parametrization of transition rates. Here we want to change this and assume rates in the form of (setting $\kT=1$)
\begin{align}
  k_{nm} = \kap \exp{\left(E_{n}-B_{nm} + \theta_{n} F_{nm} \right)} \qquad k_{mn} = \kap \exp{\left(E_{m}- B_{mn} + \theta_{m} F_{mn} \right)},
  \label{eq_SI:parametrization}
\end{align} 
where $E_{\lbrace n,m \rbrace}$ are state parameters of the states $n$ and $m$, $B_{nm}=B_{mn}$ are symmetric edge parameters, $F_{nm}= -F_{mn}$ are asymmetric edge parameters, and $\theta_{\lbrace n, m \rbrace}$ are load sharing factors that satisfy $\theta_{n} + \theta_{m} =1$. This parametrization obeys local detailed balance. Instead of the whole rate, we will now consider perturbations to only the (a)symmetric edge parameter of the input $I$. For the response of steady-state probabilities to changes in the symmetric parameter $B_{ij}$ we find  
\begin{align}
  \pd{p_n}{B_{ij}} &= \pd{p_n}{\kp} \pd{\kp}{B_{ij}} + \pd{p_n}{\km} \pd{\km}{B_{ij}} \nonumber\\
  &= -\pd{p_n}{\kp} \left( \kp - \km \frac{p_j}{p_i}  \right) \label{eq_SI:response_sym} \\ \nonumber
  & = \frac{\left[\adj(\mat{W}_i)\right]_{n, j}}{\det(\mat{W}_i)} j_{ij}.
\end{align}
For the second equality we used Eq.~(5) of the main text and the last equality was obtained by inserting Eq.~(A2). Note that the final line recovers a response relation previously derived by Aslyamov \& Esposito~\cite{aslyamov24}. Analogously we find for the asymmetric edge parameter
\begin{align}
  \pd{p_n}{F_{ij}} &=  \pd{p_n}{\kp} \left(\theta_{i} \kp + \theta_{j} \km  \frac{p_j}{p_i} \right) \nonumber \\
  &= -\frac{\left[\adj(\mat{W}_i)\right]_{n, j}}{\det(\mat{W}_i)} \left(\theta_{i} \kp p_i + \theta_{j} \km p_j   \right).
  \label{eq_SI:response_asym}
\end{align}
In case of symmetric load sharing ($\theta_{i} = \theta_{j} = 1/2$) the term in parentheses reduces to $t_{ij}/2$ and we again recover the result from Ref.~\cite{aslyamov24}. In case the load is fully carried by one of the transitions (e.g., $\theta_{i}=1$) the parentheses instead reduce to the probability flux along the load carrying transition ($J_{i\to j} = p_i k_{+I}$). Note that perturbing state parameters (or energies) $E_n$ would influence all rates leading out of the state and is thus not permitted here. However, it was found that perturbation of such kinetic properties results in a constant ratio of currents withing the network~\cite{mallory20}.

The results from Eqs.~\eqref{eq_SI:response_sym} and~\eqref{eq_SI:response_asym} extend straightforwardly to the general observables from Eq.~(9) of the main text, yielding the response relations
\begin{multline}
  \begin{bmatrix}  
  \partial_{B_{ij}} \\[.7ex]
  \partial_{F_{ij}}
  \end{bmatrix} 
  \mathcal{O}_A
  = \frac{1}{\det(\mat{W}_i)} \left\lbrace \al \sum_{m\in A} c_m \left[\adj(\mat{W}_i)\right]_{m, j} + \beta \sum_{\substack{l,m \in A}} d_{ml} k_{ml} \left[\adj(\mat{W}_i)\right]_{m, j} \right\rbrace 
  \begin{bmatrix}  
  j_{ij} \\[.7ex]
  (\theta_i \kp p_i + \theta_j \km p_j)
  \end{bmatrix}
  \\
  + \delta_{I \in A} \beta 
  \begin{bmatrix}  
  -(d_{ij} \kp p_i + d_{ji} \km p_j) \\[.7ex]
  (d_{ij}\theta_i \kp p_i - d_{ji} \theta_j \km p_j)
  \end{bmatrix}.
  \label{eq_SI:obs_response}
\end{multline}
Note that we include possible transitions through the input edge by summing over all transitions between the states of $A$ and $\delta_{I \in A}$ is unity if $i$ and $j \in A$ and zero otherwise. Equation~\eqref{eq_SI:obs_response} offers analytical expressions for the response of observables $\mathcal{O}_A$ to changes in a single rate parameter. For $\mathcal{O}_A= \lbrace j_{ij}, t_{ij} \rbrace$ and $\theta_i=\theta_j=1/2$, we once more recover the results of Ref.~\cite{aslyamov24} and augment them with additional response relations that hold for non-local perturbations. 

\section{Numerical evaluation of spanning-tree polynomials}
To conveniently calculate spanning tree polynomials we make use of the matrix-tree theorem for weighted graphs~\cite{hill66, schnakenberg76, tutte05}. First we define the weighted Laplacian matrix
\begin{align}
 \left[\bm{\mathcal{L}}\right]_{n,m} = -\left[\vec W\right]_{n,m}=  
 \begin{cases} 
   \sum_{l\neq n} k_{nl} &\text{for $n=m$} \\
   - k_{mn} &\text{otherwise,} 
 \end{cases} 
\end{align}
as minus the generator $\vec W$ of the Markovian jump process.
The matrix-tree theorem then states that the sum of all rooted spanning tree polynomials is given by
\begin{align}
  \tau_n = \det(\bm{\mathcal{L}}_{\backslash (n,n)})
\end{align}
with the reduced Laplacian matrix $\bm{\mathcal{L}}_{\backslash (n,n)}$ obtained by deleting the $n$th row and column. Similarly, one can calculate any $\tau_{n\backslash i\to j} = \det(\bm{\mathcal{L}}_{\backslash (n,n)|\kp=0})$ by evaluating the reduced matrix at $\kp = 0$; $\tau_{n\backslash j\to i}$ follows analogously. From aboves expressions one can straightforwardly deduce 
\begin{align}
  \tau_n^{/i\to j } = \left[ \det(\bm{\mathcal{L}}_{\backslash (n,n)}) - \det(\bm{\mathcal{L}}_{\backslash (n,n)|\kp=0} ) \right]_{\kp =1},
\end{align}
and similar for $\tau_n^{/j\to i}$. Note that one has to evaluate at $\kp = 1$ since, by definition, $\tau_n^{/i\to j}$ does not contain the rate associated to the transition $i \to j$. These equations offer a simple way to numerically evaluate spanning tree polynomials and offers a direct route to calculate the coefficients of the affine transformations in Eqs.~(2) and (10) through the bounds in Eqs.~(7) and (11).

\section{Illustrations}
In this section, we provide additional details on the examples we study in Figs.~1, 2, and 3 of the main text and Fig.~\ref{fig:4state_LDB} above. 
\subsection{4-state model}
\subsubsection{Figure 1}
For the 4-state model depicted in Fig.~1(a), we randomly sample transition rates $k_{nm}$ from a uniform distribution spanning the interval $0\leq k_{nm} \leq 10$. The results depicted in Fig.~1(b)-(d) were obtained for the transition rates given in Table~\ref{tab:4state}. We designate $1\leftrightarrow 2$ as input edge and assume rate $\kp \equiv k_{12}$ to be the subject of external control, taking values $k_{12}\in[0, 10^4]$. 
\begin{table}[h]
  \centering
  \caption{Transition rates for the 4-state model studied in Fig.~1 of the main text. The input edge is $I \equiv 1 \leftrightarrow 2$ and we choose to perturb rate $k_{12}$.}
  \begin{tabular}{@{\hskip 1em}cc@{\hskip 1em}|@{\hskip 1em}cc@{\hskip 1em}|@{\hskip 1em}cc@{\hskip 1em}|@{\hskip 1em}cc@{\hskip 1em}|@{\hskip 1em}cc@{\hskip 1em}}
  \hline
  \hline
  \rule{0pt}{2ex}  transition & rate $k_{nm}$&  transition & rate $k_{nm}$ & transition & rate $k_{nm}$ & transition & rate $k_{nm}$ & transition & rate $k_{nm}$ \rule[-1ex]{0pt}{0pt}\\
  \hline
  \hline
  $1 \to 2$ & $[0,10^4]$ & $2 \to 3$ & 7.2097 & $1 \to 3$ & 0.5077 & $3 \to 4$ & 3.0227 & $1 \to 4$ & 5.8359 \\
  $2 \to 1$ & 2.8911 & $3 \to 2$ & 0.2162 & $3 \to 1$ & 2.0592 & $4 \to 3$ & 6.6391 & $4\to 1$ &  3.0811 \\
  \hline
  \hline
  \end{tabular}
  \label{tab:4state}
\end{table}

\subsubsection{Figure~\ref{fig:4state_LDB}}
\label{sec:4stateLDB}
In deriving the thermodynamic bounds of Eq.~\eqref{eq_SI:TD_bounds}, rates are assumed to obey local detailed balance. This is not the case for the rates we employed in Fig.~1 (cf. Table~\ref{tab:4state}) and here we instead use rates parametrized according to Eq.~\eqref{eq_SI:parametrization}. For convenience, we take $\kap=1$, assume transitions to share the load equally ($\theta_n = \theta_m =1/2$), and set all symmetric edge parameters to zero ($B_{nm}=0$). Energies and asymmetric parameters are drawn from uniform distributions over the intervals $0 \leq E_n \leq 1$ and $-1 \leq F_{nm} \leq 1$. We perturb the transition rates of edge $I= 1\leftrightarrow 2$, by varying $F_{12} = -F_{21} \in [-1,1]$. The remaining values we used for Fig.~\ref{fig:4state_LDB}(c) are documented in Table~\ref{tab:4state_LDB}. 

\begin{table}[h]
  \centering
  \caption{State $E_n$ and asymmetric edge parameters $F_{nm}$ for the four state network studied in Fig.~\ref{fig:4state_LDB}. Transition rates that obey local detailed balance are constructed according to Eq.~\eqref{eq_SI:parametrization}.}
  \begin{tabular}{@{\hskip 1em}c@{\hskip 1em}c@{\hskip 1em}|@{\hskip 1em}c@{\hskip 1em}c@{\hskip 1em}||@{\hskip 1em}c@{\hskip 1em}c@{\hskip 1em}|@{\hskip 1em}c@{\hskip 1em}c@{\hskip 1em}}
  \hline
  \hline
  $E_1$ & 0.8734 & $E_3$ & 0.8692 & $F_{13}=-F_{31}$ &  -0.1391& $F_{34}=-F_{43}$ & -0.1953 \\
  $E_2$ & 0.9385 & $E_4$ & 0.5309 & $F_{23}=-F_{32}$  & -0.9772 &  $F_{41}=-F_{14}$ & 0.0453 \\
  \hline
  \hline
  \end{tabular}
  \label{tab:4state_LDB}
\end{table}

\subsection{Sensory adaptation---Figure~2}

\begin{figure}
  \centering
  \includegraphics{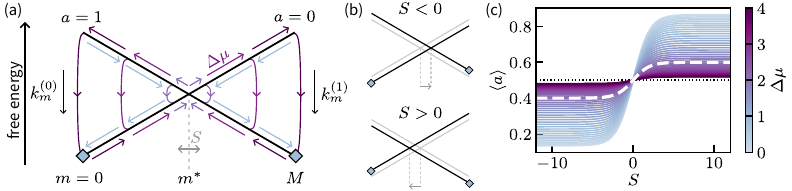}
  \caption{(a)~Linear free energy surfaces corresponding to inactive ($a=0$) and active ($a=1$) in the adaptation model of Fig.~2 of the main text. Colored arrows indicate probability currents if $\Delta \mu$ is smaller (blue) or larger than $\Delta \mu_\mathrm{c} = 2$ (purple). In the former case, the system relaxes into either of the two states indicated by blue markers, whereas the latter stabilizes dissipative cycles that grow with $\Delta \mu$ [cf.~arrow color and the colorbar in (c)]. (b) Perturbations in the external signal $S$ shifts the free energy crossover within the local environment of $m^\ast$. (c) Average activity $\langle a \rangle$ [Eq.~(A7)] as a function of the input signal $S$. Colors indicate the strength of driving $\Delta \mu$. The white dashed line corresponds to the onset of adaptation at $\Delta \mu_\mathrm{c} = 2$ and the dotted black line depicts the adapted activity level at $\langle a \rangle =0.5$.}
  \label{fig:chemo_SI}
\end{figure}

In Fig.~2, we apply our results to a simplified sensory adaptation model that is motivated by the \emph{Escherichia Coli} chemotaxis pathway~\cite{tu08, lan12, murugan17}. Only the input edge at $m=m^\ast$ depends on the external signal $S$ and we parametrize transition rates $k_m^{(1)}$ from inactive to active and $k_m^{(0)}$ from active to inactive as 
\begin{align}
  k_m^{(1)} \equiv 
  \begin{cases} 
    \tilde{\kap}_0 & (m<m^\ast) \\
    \displaystyle \frac{\kap_0}{1+\exp(-S)} & (m=m^\ast)  \\
    \kap_0 & (m>m^\ast)
  \end{cases}
  \qquad \qquad 
  k_m^{(0)} \equiv 
  \begin{cases} 
    \kap_0 & (m<m^\ast) \\
    \displaystyle \frac{\kap_0}{1+\exp(S)} & (m=m^\ast)\\
    \tilde{\kap}_0 & (m > m^\ast),
    \label{eq:rates_act}
  \end{cases}
\end{align}
respectively. For simplicity, we set $\kT =1$, choose the constant rates $\kap_0=1$, $\tilde{\kap}_0 = 10^{-3}$, and vary $S\in[-12, 12]$. 
Transitions between different $m$ along the activity axes are externally driven up the free energy landscape through affinity $\Delta \mu$ and we parametrize rates as
\begin{align}
  \om^{(0)}_+ &= \frac{\om \exp(\Delta \mu)}{1+\exp[\Delta F^{(0)}]} \qquad \om^{(0)}_- = \frac{\om}{1+\exp[-\Delta F^{(0)}]} \\
  \om^{(1)}_+ &= \frac{\om }{1+\exp[\Delta F^{(1)}]} \qquad \om^{(1)}_- = \frac{\om \exp(\Delta \mu)}{1+\exp[-\Delta F^{(1)}]},
  \label{eq:rates_meth}
\end{align}
where the free energy differences between consecutive methylation levels are taken as $\Delta F^{(0)} = - \Delta F^{(1)} = 2$ and attempt rate $\om = 2$ defines the transition timescale. The corresponding effective free energy landscape is depicted in Fig.~\ref{fig:chemo_SI}(a). The free energy surfaces are linear with respect to the methylation level $m$, with offset and slope that depend on $a$ so that both surfaces intersect at $m^\ast$. For small driving $\Delta \mu < \Delta \mu_\mathrm{c}$ (for the chosen parameters $\Delta \mu_c =2$, see~\cite{sartori15}) the system relaxes into either $(m=0, a=0)$ or $(m=M, a=1)$ as indicated by the blue arrows and markers. Which of the two is favored depends on the input signal $S$. Since its influence is restricted to the input edge at $m=m^\ast$, changing $S$ effectively shifts the free energy crossover within the local environment $m^\ast \pm 1$ to bias relaxation into either of the two states, while the other becomes metastable [Fig.~\ref{fig:chemo_SI}(b)]. The situation depicted in Fig.~\ref{fig:chemo_SI}(a) corresponds to $S=0$ where relaxation into both states is equally likely. For $S<0$ ($S>0$), relaxation into the passive $a=0$ (active $a=1$) state becomes energetically favorable and renders the output activity highly sensitive to $S$ [see Fig.~\ref{fig:chemo_SI}(c)].

Conversely, for strong driving $\Delta \mu > \Delta \mu_\mathrm{c}$ a stable cycle around the adapted methylation level ($m^\ast$ for $S=0$) emerges through driving transitions against the free energy gradient [purple arrows in Fig.~\ref{fig:chemo_SI}(a)]. This mechanism assures a robust activity output that is sustained irrespective of external influences [Fig.~\ref{fig:chemo_SI}(c)]. The adaptation error, i.e., deviations to the dotted line in Fig.~\ref{fig:chemo_SI}(c), reduces upon increasing $\Delta \mu$. An increase in $\Delta \mu$ results in cycles that traverse through more levels before switching the surfaces, which are ultimately constrained and stabilized along the system's boundary [color gradient in the purple arrows of Fig.~\ref{fig:chemo_SI}(a)].

\subsection{Stochastic resetting in calmodulin folding---Figure~3}

We also illustrate our results for unidirectional input edges by considering the experimentally inferred folding network of calmodulin protein~\cite{stigler11}, which we augment by an unidirectional resetting edge. The transition rate through the resetting edge is externally controlled, taking values $r \in [0, 10^4]$ in Figs.~3(b) and (d). The remaining rates are taken from the Supplemental Material of Ref.~\cite{bebon23} (see Table~\ref{tab:calmodulin}), where the rates for a pretension of $9\,\mathrm{pN}$ (see Supplemental Material of Ref.~\cite{stigler11}) are adjusted within the experimental error bars to exactly satisfy detailed balance. 

For completeness, the bounds on the entropy production rate depicted in Fig.~3(d) are obtained from Eq.~(11) of the main text as
\begin{align}
  \mathcal{B}_{\sig_\mathrm{rev}}^+ = \sum_{n<m \in \mathcal{N}\backslash \lbrace I \rbrace} \left(\mathcal{B}_n^+ k_{nm} - \mathcal{B}_m^+ k_{mn} \right) \ln\left( \frac{k_{nm}}{k_{mn}}\right)
\end{align}
and $\mathcal{U}_{\sig_\mathrm{rev}}^+ = 0$ follows from the vanishing currents as the equilibrium limit ($r\to 0$) is approached.

\begin{table}[h]
  \centering
  \caption{Experimentally determined transition rates for the folding dynamics of a single calmodulin protein subjected to a pretension of $9\,\mathrm{pN}$~\cite{stigler11}. To account for experimental uncertainties that lead to slight deviations from detailed balance, we take the modified rates from the Supplemental Material of Ref.~\cite{bebon23}.}
  \begin{tabular}{@{\hskip 1em}cc@{\hskip 2em}|@{\hskip 2em}cc@{\hskip 2em}|@{\hskip 2em}cc@{\hskip 1em}}
  \hline
  \hline
  \rule{0pt}{2ex} transition & rate $k_{nm}$ & transition & rate $k_{nm}$ & transition & rate $k_{nm}$ \rule[-1ex]{0pt}{0pt}\\
  \hline
  \hline
  $1 \to 2$ & 5.997 & $1 \to 4$ & 13.439 & $1 \to 5$ & 15.330 \\
  $2 \to 1$ & 0.774 & $4 \to 1$ & 127.968 & $5 \to 1$ & 0.121 \\
  \hline
  \hline
  $5 \to 6$ & 3.749 & $2 \to 3$ & 1514.820 & $2 \to 6$ & 13.441 \\
  $6 \to 5$ & 13.326 & $3 \to 2$ & 53.066 & $6 \to 2$ & 2.922 \\
  \hline
  \hline
  \end{tabular}
  \label{tab:calmodulin}
\end{table}

\end{document}